# Optimal model of semi-infinite graphene for *ab initio* calculations of reactions at graphene edges by the example of zigzag edge reconstruction

Yulia G. Polynskaya[a] [*], Irina V. Lebedeva[b,c], Andrey A. Knizhnik[a,d], and Andrey M. Popov[e]

[a] Kintech Lab Ltd., 3rd Khoroshevskaya Street 12, Moscow 123298, Russia.
[b] CIC nanoGUNE BRTA, Avenida de Tolosa 76, San Sebastian 20018, Spain.
[c] Catalan Institute of Nanoscience and Nanotechnology - ICN2, CSIC and BIST, Campus UAB, Bellaterra 08193, Spain
[d] National Research Centre "Kurchatov Institute", Kurchatov Square 1, Moscow 123182, Russia.
[e] Institute for Spectroscopy of Russian Academy of Sciences, Fizicheskaya Street 5, Troitsk, Moscow 108840, Russia.

[*] Corresponding author.

*E-mail addresses:* yupol@kintechlab.com (Yu. G. Polynskaya), liv_ira@hotmail.com (I. V. Lebedeva), popov-isan@mail.ru (A. M. Popov), knizhnik@kintechlab.com (A. A. Knizhnik)

**Abstract**

We investigate how parameters of the model of semi-infinite graphene based on a graphene nanoribbon under periodic boundary conditions affect the accuracy of *ab initio* calculations of reactions at graphene edges by the example of the first stage of reconstruction of zigzag graphene edges, formation of a pentagon-heptagon pair. It is shown that to converge properly the results, the nanoribbon should consist of at least 6 zigzag rows and periodic images of the pair along the nanoribbon axis should be separated by at least 6 hexagons. The converged reaction energy and activation barrier for formation of an isolated pentagon-heptagon pair are found to be –0.15 eV and 1.61 eV, respectively. It is also revealed that such defects reduce the graphene edge magnetization only locally but ordering of spins at opposite nanoribbon edges switches from the antiparallel (antiferromagnetic) to parallel one (ferromagnetic) upon increasing the defect density.

# 1. Introduction

In the last two decades, graphene has been a subject of intensive studies because of its great potential for nanoelectronics [1,2]. One of the most important features of this nanomaterial, which makes it adequate for diverse applications, is that its properties can be tuned in wide ranges by size, shape and atomistic structure of the layer, in particular by structure of the edges and presence of defects (see, for example, reviews [3–5]). Graphene-based nanostructures, such as graphene nanoribbons (GNRs) [6,7], quantum dots [7,8] and nanojunctions [9,10] exhibit properties very different from those of bulk graphene and are extremely sensitive to the edge structure because of the high edge to bulk ratio. The effect of edge structure on mechanical [11–15], electronic [7,16 – 28], magnetic [13,14,16,27,29,30], chemical [14,21,27,29,31] and thermal [32] properties of graphene has been demonstrated. Chemical modification of graphene edges by functional groups also leads to changes of the properties listed [14,19,23–27,29,31–36]. Since structural modification of graphene edges can be used to tune their properties, significant efforts have been made recently to investigate structure [14,15,18,19,21,23,24,27,30,31,37–45], energetics [12–15,21,22,27,29,31,44–47] and structural rearrangements [12,13,21,22,38,40,42,48] of graphene edges (see also [5] for review).

Density functional theory (DFT) calculations play a crucial role in investigations of graphene edges [12–16,21–27,29,31,44–49]. However, since the size of the atomistic model that can be studied by DFT calculations is normally limited to hundreds – thousands of atoms, the adequacy of these calculations relies on the convergence of calculated values of physical quantities with respect to geometrical parameters of the model. In the present paper we investigate the effect of the parameters of the model of a semi-infinite graphene layer by the example of the reconstruction of zigzag graphene edges [14,15,21,27,31,43–47]. Note that a surface reconstruction is a relatively common phenomenon for nanowires and nanomembranes [50 – 52] in which lowering the system symmetry is energetically favourable. In the case of zigzag edge reconstruction, the number of dangling bonds at zigzag edges is reduced via transformation of pairs of hexagons (Fig. 1a) into 57 pairs (Fig. 1b) and formation of triple bonds. As a result, the energy of zigzag edges is decreased [14,15,21,27,31,43–47]. Although reconstructed zigzag edges are more stable thermodynamically than other pristine graphene edges decreased [14,15,21,31,44–47], they are not the most abundant ones in the experiments [38–42]. As an explanation, it has been proposed that fast chemical etching favors the structure of unreconstructed zigzag edges [37,39,42,44,53,54] and that it is likely that there is a significant barrier for zigzag edge reconstruction [12,13,22,40]. There is, nevertheless, a considerable scatter in activation barriers and reaction energies reported. For formation of an isolated pentagon-heptagon (57) pair (see Fig. 2c), the values of 1.12 eV [12] and 1.61 eV [22], 1.80 eV [13] were obtained via DFT calculations for zigzag graphene nanoribbons (ZGNRs) and 2.41 eV for the zigzag edge of a graphene flake [48] (see also [5] for review). The reaction energies varied from almost -1.0 [12] to +0.2 [13]. Here we intend to find the origin of these discrepancies.

To get adequate results for zigzag edge reconstruction by DFT calculations, an appropriate model of a semi-infinite graphene layer should be chosen. In most of the calculations [12,13,22], ZGNRs under periodic boundary conditions along the ZGNR axis were used as such models. However, the ZGNR width, simulation cell size along the ZGNR axis and conditions imposed on the second edge were not the same and could lead to the observed discrepancies in the results. To study the energetics of formation of an isolated defect at the edge of a semi-infinite graphene layer, it is important to exclude the interaction of periodic images of the defect, i.e. the simulation cell should be sufficiently large (large *L* in Fig. 2a). The GNR should be sufficiently wide to exclude the effect of the second edge (large *n* in Fig. 2a). Also it is sometimes assumed that hydrogen termination of the second edge (Figs. 2a and b) or its fixation [12,13] to imitate the structure of bulk graphene (*f* in Fig. 2a) can speed up the convergence of the results upon increasing the GNR width. In the present paper, we investigate the influence of these factors on the energetic parameters of formation of a 57 pair on the ZGNR edge (Fig. 2c) to come up with the adequate model for studies of reactions at graphene edges. This model can be later applied for DFT studies of formation of subsequent 57 pairs for complete zigzag edge reconstruction as well as other phenomena at graphene edges such as edge etching [42,44,54,55], formation of atomic chains [38,56,57], migration of atoms [22,37,53], graphene growth [58–60], chemical absorption [14,21,25,27,29,31–36], etc. As a result, we also obtain the converged reaction energy and activation barrier for formation of an isolated 57 pair at the zigzag graphene edge.

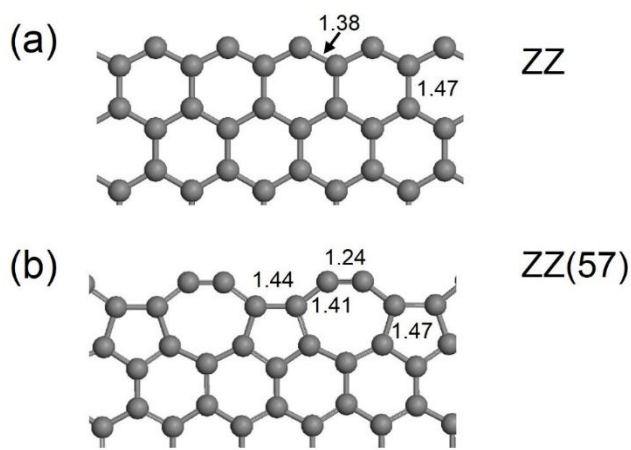

**Fig 1.** Geometries of graphene edges: (a) zigzag (ZZ) and (b) reconstructed zigzag (ZZ(57)). Calculated bond lengths in Å are indicated.

It should be noted that because of the presence of dangling bonds, localized edge states are formed at zigzag edges [17,28]. These states give rise to flat bands at the Fermi level resulting in magnetic instability, which is resolved through ordering of spins at the edge atoms [12–14,16,24,25,27,29,30,49]. The ordering of spins in the ground state is parallel, i.e. ferromagnetic, at each edge but antiparallel, i.e. antiferromagnetic, between the opposite edges if a ZGNR is considered. Upon reconstruction, however, unpaired electrons form triple bonds and the edge becomes non-magnetic in the ground state (similar to the case of armchair edges) [12,13,27]. In the present paper, we consider how the magnetism of zigzag graphene edges changes during the formation of 57 pairs and depends on the density of these defects. These results have implications for development of graphene-based spintronic devices [30,49,61,62].

The paper is organized as follows. First we describe the methods used. Then we present our results on the dependence of the relative energy of the reconstructed zigzag edge on the parameters of the ZGNR model, the magnetization change during the formation of a 57 pair and the effect of the geometrical parameters of the model on the energetics of this process. Finally we summarize our conclusions.

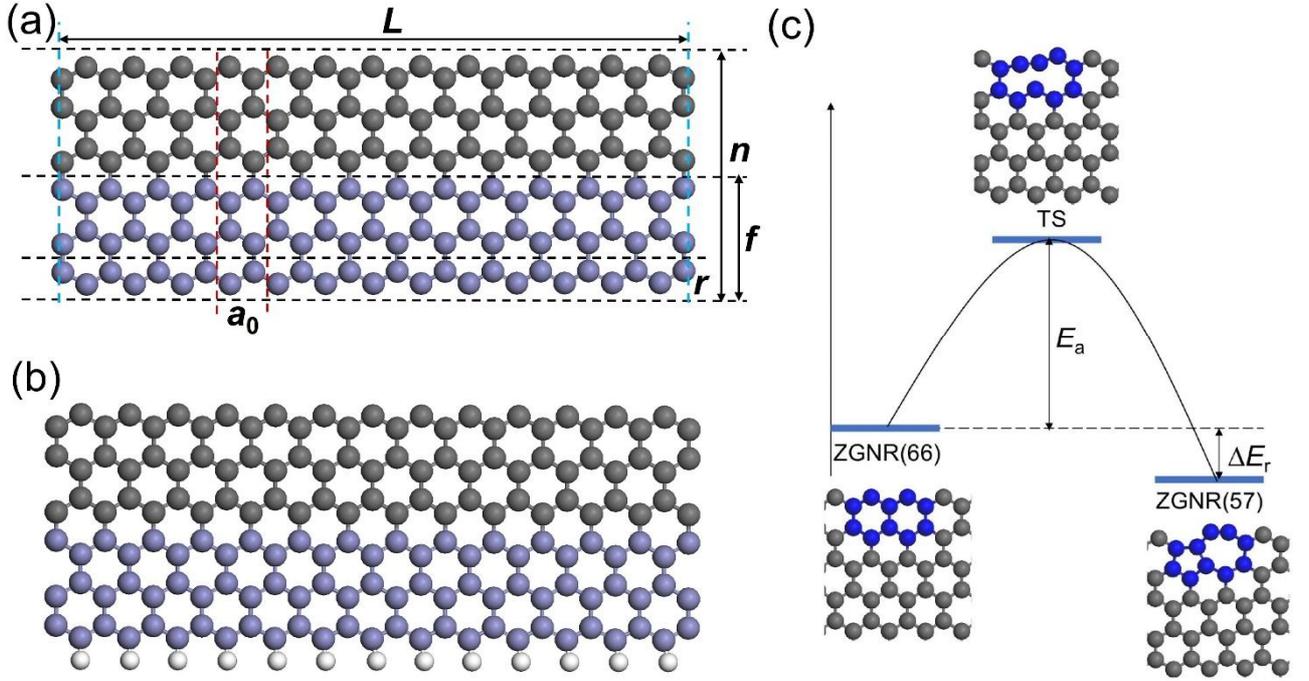

**Fig. 2**. (a) and (b) Atomistic structures of zigzag graphene nanoribbons (ZGNRs) used as models of semi-infinite graphene with the zigzag edge. The edge opposite to the one being reconstructed is terminated by carbon (a) or hydrogen (b). The elementary unit cell of the pristine ZGNR of length $a_0$ is shown by vertical red dashed lines. The cell length $L$ is indicated by vertical cyan dashed lines. One zigzag row, $r$, is shown by horizontal dashed lines. Carbon atoms of free and fixed zigzag rows are coloured in grey and purple, respectively. The total number $n$ of zigzag rows and the number $f$ of fixed zigzag rows at the edge opposite to the one being reconstructed are also indicated. Interatomic distances between the fixed atoms are the same as in bulk graphene. Hydrogen atoms are coloured in white. (c) A scheme of formation of a pentagon-heptagon (57) pair at the zigzag graphene edge. The reaction energy, $\Delta E_r$, and activation barrier, $E_a$, are shown. ZGNR(66) is the pristine ZGNR, ZGNR(57) is the ZGNR with a 57 pair, TS is the transition state for formation of the 57 pair. Atoms of rings involved in the reaction are coloured in blue.

## 2. Computational details

Structure optimization and energy calculations based on the spin-polarized DFT approach were carried out using the PBE (Perdew-Burke-Ernzerhof) functional [63] as implemented in the VASP code [64]. The interaction of valence and core electrons was described using the projector-augmented wave method (PAW) [65]. The maximal kinetic energy of the plane wave basis set was 500 eV. The Gaussian smearing method with the width of 0.05 eV was used. Self-consistent iterations were performed till the tolerance of $10^{-8}$ was reached. ZGNRs were considered as models of a semi-infinite graphene layer (Fig. 2). Correspondingly, the unit cell length of ZGNRs was taken equal to the lattice constant of graphene $a_0$ (2.466 Å according to our calculations). In the cases when atoms at the edge opposite to the one being reconstructed were fixed, interatomic distances between these atoms were the same as in bulk graphene. To stimulate convergence to a ferromagnetic or antiferromagnetic state, initial spins with the parallel or antiparallel ordering at the opposite ZGNR edges, respectively, were set at the edge atoms. Periodic boundary conditions were applied to simulation cells of different length along the ZGNR axis and width across the ZGNR. The vacuum gap of at least 10 Å across the ZGNR and perpendicular to ZGNR plane was introduced to minimize the interaction between periodic images of the ZGNR. The Brillouin-zone integration was performed using a Monkhorst-Pack $k$ point sampling [66]. $36a_0/L$ k-points were used for the simulation cell of length $L$ along the ZGNR axis. The positions of atoms in pristine ZGNRs and ZGNRs with 57 pairs were optimized until forces on the atoms became less than 0.003 eV/Å.

The pathway of formation of the first 57 pair at the ZGNR edge, i.e. the first step of the edge reconstruction, was investigated using the nudged elastic band (NEB) method [67,68]. 6 images between the initial and final states were considered. To reduce the computational effort, the images were geometrically optimized for the maximal kinetic energy of the plane wave basis set of 400 eV, the convergence criterion for self-consistent iterations of $10^{-4}$ and smearing width of 0.2 eV. The optimization was performed via damped molecular dynamics until the maximal residual force of 0.03 eV/Å. After optimization, a single-point energy calculation was performed to estimate the energy of the transition state with a higher accuracy using the maximal kinetic energy of the plane wave basis set of 500 eV, the convergence criterion for self-consistent iterations of $10^{-8}$ and smearing width of 0.05 eV.

The reaction energy ($\Delta E_r$) was calculated as the difference between the energy $E_{57}$ of the ZGNR with a single 57 pair per the simulation cell and energy $E_{66}$ of the ZGNR with pristine zigzag edges (Fig. 2c): $\Delta E_r = E_{57} - E_{66}$. The activation barrier was computed as $E_a = E_{TS} - E_{66}$, where $E_{TS}$ is the energy of the transition state for formation of the 57 pair. The difference in the energies of the reconstructed and unreconstructed zigzag edges was found as $\Delta \varepsilon_r = \Delta E_r/(2a_0)$ from the results for the simulation cell of length $2a_0$.

The VASPKIT [69] tool was applied to extract the spin density from the VASP output. The spin maps were visualized using VESTA (Visualisation for Electronic Structural Analysis) [70]. Since the spin density at each zigzag edge is mostly concentrated at atoms of one sublattice and extends far from the edge atoms in the direction perpendicular to the edge (see Fig. 4 below), we used a relatively large Wigner-Seitz radius of $a_0/2$ to estimate magnetic moments of individual atoms. Note that the sum of the magnetic moments of atoms estimated in this way almost reproduces the total magnetic moment of the simulation cell for the ZGNR in the ferromagnetic state with the parallel ordering of spins at the opposite ZGNR edges.

## 3. Results and discussion
### a) Relative energy of the reconstructed zigzag edge

Thermodynamics of the edge reconstruction is determined by the relative energy of the reconstructed zigzag edge. Therefore, we first consider how the parameters of the model ZGNR affect the relative energy of the reconstructed zigzag edge with respect to the unreconstructed one (Figs. 1b and a, respectively). The effect of the following parameters of the ZGNR is studied: (1) the total number of zigzag rows ($n$ in Fig. 2a), (2) the number of zigzag rows at the edge opposite to the one being reconstructed that are fixed with interatomic distances between the fixed atoms equal to those in bulk graphene ($f$ in Fig. 2a) and (3) the way the opposite edge is terminated (by carbon or hydrogen, Figs. 2a and b, respectively). The fixation of atoms at the opposite edge has been used in literature to stimulate the convergence to the geometry of the edge of a semi-infinite graphene layer [12,13]. The hydrogen termination can be applied to saturate dangling bonds at the opposite edge and stabilize the ZGNR structure. To compute the edge energies, we consider the minimal simulation cell with only two hexagons or one 57 pair along the ZGNR axis, i.e. the length of the simulation cell is $L = 2a_0$. Note that the energy difference $\Delta\varepsilon_r$ for the reconstructed and unreconstructed zigzag edges is related to the reaction energy $\Delta E_r$ for the complete edge reconstruction as $\Delta E_r = 2a_0\Delta\varepsilon_r$.

The dependence of the relative energy $\Delta\varepsilon_r$ of the reconstructed zigzag edge with respect to the unreconstructed one on the total number $n$ of zigzag rows in the range of $n = 4 - 12$ is presented in Fig. 3a. It is seen that for the fully free $n$-ZGNRs ($f = 0$), the relative energy of the reconstructed zigzag edge decreases monotonically upon increasing the number $n$ of zigzag rows. Such a dependence of the relative energy $\Delta\varepsilon_r$ of the reconstructed edge on the ZGNR width can be attributed to the quantum confinement effects [16,28,71]. Similar monotonic dependences on the GNR width were also observed in Ref. [22] for the formation energy of a single 57 pair at the unreconstructed zigzag edge and in Refs. [14,15,21,24] for energies of diverse graphene edges except the armchair one. Oscillations of the energy of the armchair edge upon changing the GNR width are correlated with oscillations of the band gap and should be also caused by the quantum confinement effects [14,15].

The variation of the relative energy $\Delta\varepsilon_r$ of the reconstructed edge in the range of considered ZGNR widths is, nevertheless, rather small (Fig. 3). For the 4-ZGNR, the deviation from the result for the widest considered 12-ZGNR is only 0.01 eV/Å, i.e. 0.05 eV per 57 pair. For the 6-ZGNR, the deviation is already within 0.006 eV/Å, i.e. 0.03 eV per 57 pair. These deviations are close to the so-called "chemical accuracy", i.e. the accuracy required to make realistic chemical predictions and normally assumed in literature to be 1 – 2 kcal/mol [72,73]. Therefore, for accurate calculations of the relative energy of graphene edges, it is sufficient to consider narrow ZGNRs consisting of 4 or more zigzag rows. This conclusion is consistent with previous observation that absolute energies of diverse graphene edges converge fast upon changing the GNR width [14,15,21,24]. The converged relative energy $\Delta\varepsilon_r$ of the reconstructed zigzag edge of –0.20 eV/Å according to our calculations is in agreement with the previously reported DFT values (Table 1).

The dependence of the relative energy $\Delta\varepsilon_r$ of the reconstructed edge obtained for the fully free ZGNRs with the hydrogen termination of the edge opposite to the one where the reconstruction takes place (Fig. 3b) is very similar to that for the case of the carbon-terminated opposite edge (Fig. 3a). The deviation $\delta\varepsilon_{r,12}$ of the relative energy $\Delta\varepsilon_r$ of the reconstructed edge from the result for the widest 12-ZGNR is different for the two terminations of the opposite edge only for very narrow ZGNRs with 6 or less zigzag rows (Fig. 3c). For the larger ZGNRs, the edge termination does not influence the dependence of the $\delta\varepsilon_{r,12}$ on the ZGNR width. Therefore, the hydrogen termination improves the convergence of $\Delta\varepsilon_r$ only for very narrow ZGNRs consisting of less than 6 zigzag rows. It should be noted that there is a small difference in the values of $\Delta\varepsilon_r$ for the 12-ZGNR in the cases of the hydrogen and carbon-terminated opposite edge of about 1.5 meV/Å. We attribute this difference to the charge balance. The edge states at the unreconstructed edge of the considered ZGNRs opposite to the one where the 57 pairs are formed correspond to a particular spin polarization both in the carbon and hydrogen-terminated ZGNRs [12–14,16,24,25,27,29,30,49]. However, the total magnetic moment of this edge is much higher for the ZGNRs with the carbon-terminated edge as compared to the hydrogen-terminated one. This means that the occupation of the edge states at the edge being reconstructed with a different spin polarization is smaller for the carbon-terminated ZGNRs (see the band structures for ZGNRs with different edges in Refs. [14,21,27,29,31]) and the reaction energy for the reconstruction is slightly affected.

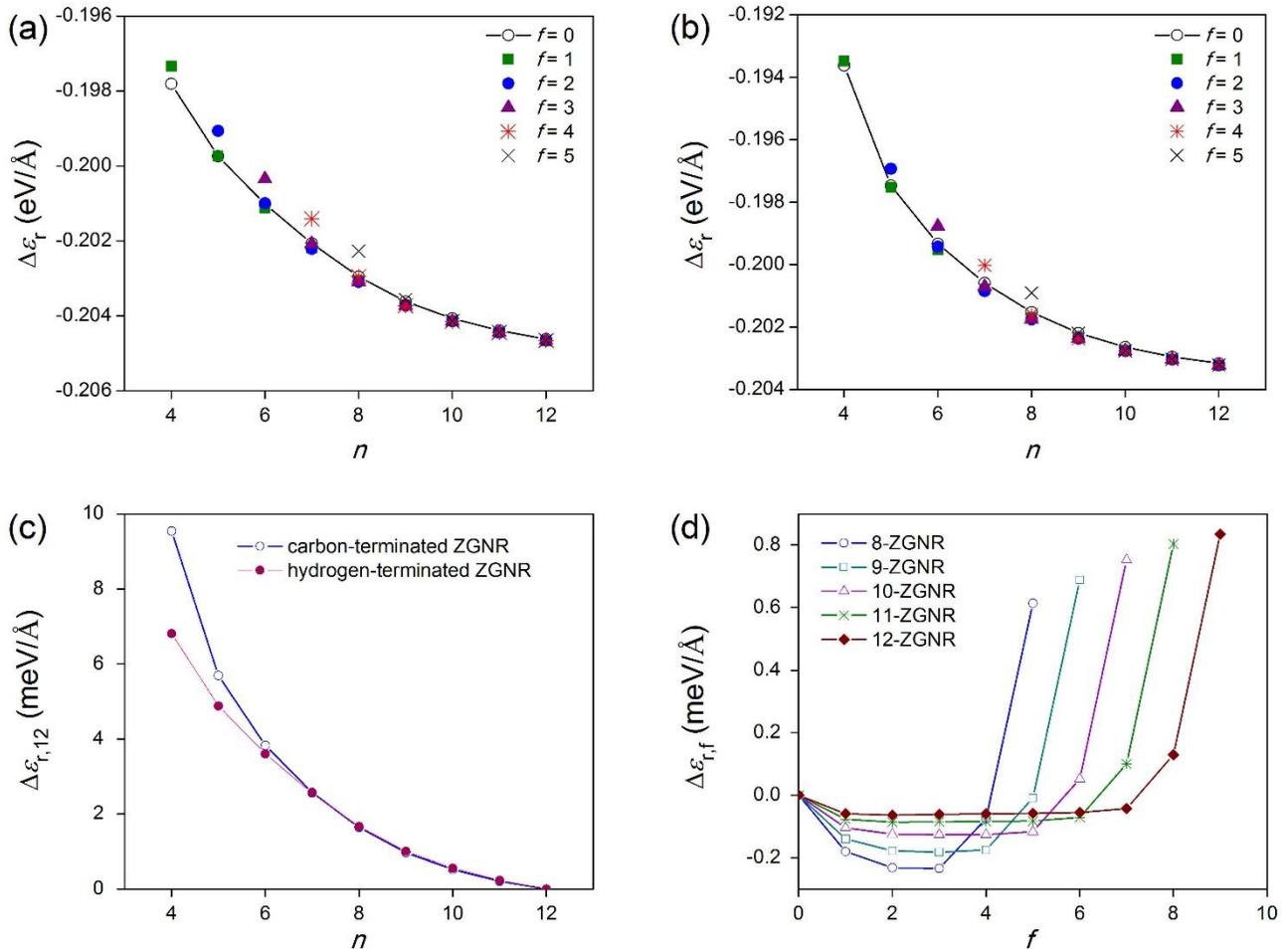

**Fig. 3.** (a,b) Calculated relative energy $\Delta\varepsilon_r$ (in eV/Å) of the reconstructed zigzag edge with respect to the unreconstructed one computed using (a) carbon-terminated and (b) hydrogen-terminated zigzag graphene nanoribbons (Figs. 1a and b, respectively) with a different total number $n$ of zigzag rows and different numbers $f$ of fixed zigzag rows at the edge opposite to the one being reconstructed: no zigzag rows fixed (solid line with open circles), 1 (green squares), 2 (blue circles), 3 (purple triangles), 4 (red stars) and 5 (black crosses) zigzag rows fixed. The solid line is shown for the case without fixed zigzag rows to guide the eye. Interatomic distances between fixed atoms are the same as in bulk graphene. (c) The deviations of the relative energy $\delta\varepsilon_{r,12}$ (in meV/Å) of the reconstructed zigzag edge from the result for the nanoribbon consisting of 12 zigzag rows for the fully free carbon (open circles) and hydrogen-terminated (filled circles) nanoribbons with a different number $n$ of zigzag rows. (d) The change $\delta\varepsilon_{r,f}$ (in meV/Å) in the relative energy $\Delta\varepsilon_r$ of the reconstructed zigzag edge upon fixing $f$ zigzag rows with respect to the fully free carbon-terminated nanoribbon with a different total number $n$ of zigzag rows: (circles) 8, (squares) 9, (triangles) 10, (crosses) 11 and (diamonds) 12.

**Table 1.** DFT data on the relative energy $\Delta\varepsilon_r$ of the reconstructed zigzag edge with respect to the unreconstructed one.

| Ref. | DFT approach | $\Delta\varepsilon_r$, eV/Å |
|---|---|---|
| This work | PBE | −0.20 |
| [46] | PBE | −0.17 |
| [31] | PBE | −0.18 |
| [47] | PBE | −0.20 |
| [43] | PBE | −0.15 |
| [21] | PBE[a] | −0.35 |
| [27] | PW91[b] | −0.24 |
| [15] | GGA[c] | −0.20 |
| [45] | LDA[d] | −0.25 |
| [14] | LDA | −0.24 |
| [44] | LDA[a] | −0.37 |

[a] Spin polarization not taken into account.
[b] Perdew-Wang 91 functional.
[c] Generalized gradient approximation (functional not specified).
[d] Local density approximation.

The fixation of atoms at the edge opposite to the one where the reconstruction takes place in such a way that interatomic distances between the fixed atoms are the same as in graphene does not improve the convergence at all (Fig. 3a and b). Upon fixing, the relative energy $\Delta\varepsilon_r$ of the reconstructed zigzag edge with respect to the unreconstructed one is reduced by 0.3 meV/Å, i.e. 1 meV per 57 pair, at most. The fact that $\Delta\varepsilon_r$ is determined mostly by the total ZGNR width and not by the width of the free ZGNR part ($n$ and not $n − f$ in Fig. 2a) cannot be explained by the structural relaxation as proposed in Ref. [22]. It is more likely that the deviation of $\Delta\varepsilon_r$ from the value for semi-infinite graphene is the result of the quantum confinement effects [16,28,71]. Although fixing some zigzag rows does not help to speed up the convergence with respect to the ZGNR width, it can be used to reduce the number of degrees of freedom considered for the geometry optimization and transition state search. It should be noted, however, that leaving only 3 free zigzag rows ($n − f$ = 3) leads to an increase in the relative energy $\Delta\varepsilon_r$ of the reconstructed zigzag edge by up to 1 meV/Å, i.e.

5 meV per 57 pair. The value of $\Delta\varepsilon_r$ stops to depend on the number $f$ of the rows fixed once there are 5 or more free rows (Fig. 3d). In the latter case, $\Delta\varepsilon_r$ is slightly smaller than that for the free ZGNR of the same width.

To summarize, such computational tricks as the fixation of atoms at the edge opposite to the one being reconstructed or changing the way that edge is terminated in addition to being unrealistic, hardly help to reduce the deviation of the energy difference between the reconstructed and unreconstructed zigzag edges in the ZGNRs from the converged value expected for semi-infinite graphene. Therefore, below we investigate the first stage of the reconstruction, formation of a 57 pair, for fully free carbon-terminated ZGNRs.

### b) Edge magnetism during the formation of a 57 pair

As we show in the following subsection, the proper account of the edge magnetization is extremely important for accurate calculations of the zigzag edge reconstruction. Therefore, before consideration of the energetics of formation of the first 57 pair, let us discuss how this process affects the zigzag edge magnetization [12–14,16,24,25,27,29,30,49]. We show in Fig. 4 how the spin density changes upon formation of the first 57 pair in the 6-ZGNR for the simulation cell of length $L = 14a_0$. In agreement with previous papers [12–14,16,24,25,27,29,30,49], we observe that edge atoms of the unreconstructed zigzag edges carry significant spins. The magnetic moments of these atoms reach almost one Bohr magneton, $\mu_B$ (Fig. 4a). The spins are ordered in the antiferromagnetic manner at the opposite ZGNR edges. According to our calculations for the 6 and 12-ZGNRs, the ferromagnetic state is higher in energy than the antiferromagnetic one by about 6 meV/Å and 2 meV/Å, i.e. 16 meV per and 4 meV per hexagon, respectively. These results are consistent with the previously reported values [12,13,16,24,25,27]. In the transition state for formation of the first 57 pair, magnetic moments of the edge atoms belonging to the defect get already considerably reduced (Fig. 4b). In the final state (Fig. 4c), the atoms of the 57 pair that were two-coordinated in the initial structure have magnetic moments of about $0.2\mu_B$. Magnetic moments of the rest of the atoms involved in the defect are within $0.05\mu_B$. Magnetic moments of the edge atoms close to the 57 pair are only slightly decreased compared to the unreconstructed edge (within $0.1\mu_B$), while magnetic moments of atoms beyond one hexagon from each side of the 57 pair are virtually not affected at all by the defect formation. Our results are thus consistent with calculations of Ref. [48] for a graphene flake and demonstrate the edge atoms of the isolated 57 pair retain a fraction of their magnetic moments as compared to the unreconstructed edge, different from the case of the completely reconstructed edge, which is not magnetic at all [12,13,27].

According to our calculations, the antiferromagnetic ordering of spins at the opposite ZGNR edges is preserved upon the formation of a 57 pair in long simulation cells (Fig. 4). The formation of isolated 57 pairs reduces the edge magnetization only locally, at the atoms involved in the defect and thus it can be expected that spin transport in spintronic devices based on graphene nanoribbons should not be affected strongly as long as the density of such defects is low [30,48,61,62]. In small simulation cells of length $L = (4 - 8)a_0$, however, we find that the ferromagnetic ordering of spins at the opposite ZGNR edges becomes more energetically favourable than the antiferromagnetic one upon the formation of the 57 pair (Fig. 5). In other words, switching from the antiferromagnetic to ferromagnetic state should occur upon increasing the density of 57 pairs or decreasing the distance between them. The relative energy of different magnetic states and the density of 57 pairs at which the transition should occur depend on the ZGNR width and they are discussed in the following subsection. This phenomenon can be used to tune spin transport in nanoribbons considered for applications in spintronic devices [30,48,61,62]. The creation of the properly arranged 57 pairs in such nanoribbons can stabilize the ferromagnetic state with different conductivities for spin up and spin down [13,24,25]. Note that the effect of 57 pairs on the edge magnetization is distinct from the case of the previously considered defects at the zigzag edge such as vacancies [27,74] and substitutional dopants [74], which were shown to suppress the edge magnetization completely upon increasing their concentration up to 1 defect per about 10 Å. In the case of 57 pairs, the magnetization is suppressed only at the atoms within the defect, while the other edge atoms maintain their magnetic moments up to very high defect concentrations.

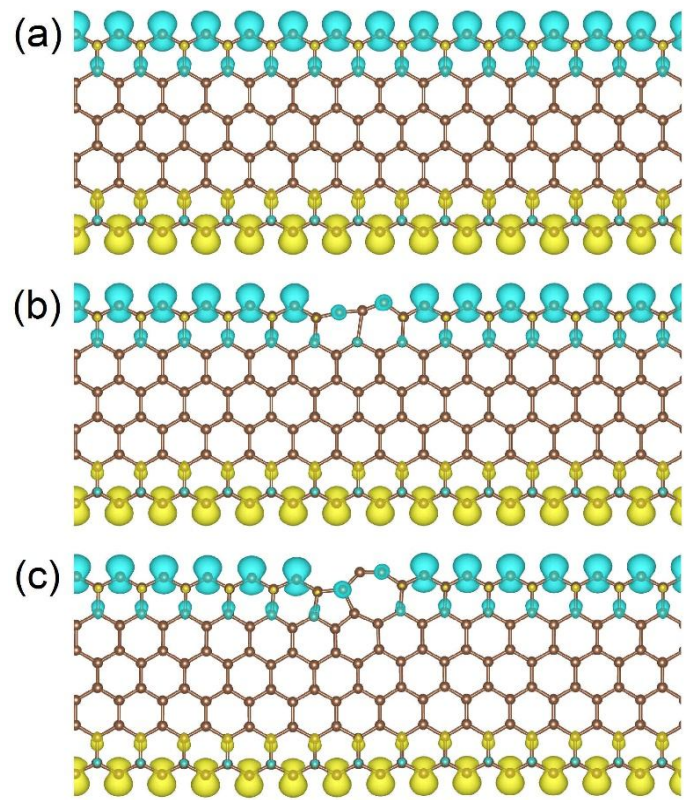

**Fig. 4.** Calculated atomistic structure and spin maps (isosurface 0.01 a. u.) during the formation of a pentagon-heptagon (57) pair at the edge of the fully free and carbon-terminated zigzag graphene nanoribbon consisting of 6 zigzag rows in the simulation cell of length $L = 14a_0$, where $a_0$ is the lattice constant of graphene: (a) the pristine zigzag graphene nanoribbon, (b) the transition state for formation of the 57 pair and (c) the nanoribbon with the 57 pair.

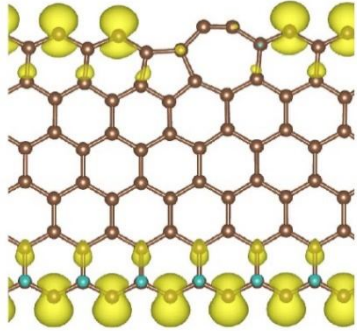

**Fig. 5.** Calculated atomistic structure and spin map (isosurface 0.01 a. u.) for the fully free and carbon-terminated zigzag graphene nanoribbon with a pentagon-heptagon (57) pair consisting of 6 zigzag rows in the simulation cell of length $L = 6a_0$, where $a_0$ is the lattice constant of graphene.

### 3.3 Energetics of the formation of a 57 pair

Let us now consider how the geometrical parameters of the model, namely the ZGNR width and length of the simulation cell along the ZGNR axis, influence the energetics of the formation of a 57 pair at the unreconstructed zigzag edge. In Fig. 6, we present the computed dependences of the reaction energy $\Delta E_r$ on the ZGNR width for lengths of the simulation cell of $L = 4a_0$, $6a_0$ and $8a_0$. The results for the final states of the ZGNR with a 57 pair with the antiferromagnetic and ferromagnetic ordering of spins at the opposite ZGNR edges are shown.

It is seen from Fig. 6a that for each magnetic state of the ZGNR with a 57 pair and length of the simulation cell considered, the reaction energy decreases monotonically upon increasing the ZGNR width starting from the 8-ZGNR. Such a dependence is in agreement with the calculations of Ref. [22] for the simulation cell of length $L = 8a_0$ and can be attributed to the quantum confinement effects [16,28,71], as discussed in the previous section for the complete zigzag edge reconstruction. For the narrow ZGNRs consisting of 7 and fewer zigzag rows, however, the dependence is no longer monotonic. It is seen that the reaction energies computed for the 4-ZGNR deviate from the results for the wider ZGNRs with the same magnetic state of the ZGNR with a 57 pair and equal lengths of the simulation cell by up to 0.1 eV, while the data for the 6-ZGNR show much smaller deviations within 0.06 eV, close to the chemical accuracy [72,73]. The same difference of the reaction energies was reported for the 6- and 12-ZGNRs in Ref. [22]. Thus, for accurate modeling of structural rearrangements at zigzag graphene edges, it can be recommended to consider ZGNRs consisting of at least 6 zigzag rows.

The difference in the reaction energies $\delta E_{r,s}$ for the final states of the ZGNR with a 57 pair with the ferromagnetic and antiferromagnetic spin ordering has a minimum for narrow ZGNRs and decreases in magnitude monotonically for wide ZGNRs when the simulation cells of length $L = (4 – 8)a_0$ are considered (Fig. 6b). For the 4-ZGNR, the antiferromagnetic final state is always energetically favourable over the ferromagnetic one. For the 6-ZGNR, the ferromagnetic final state is lower in energy than the antiferromagnetic one for the simulation cells of length $L = 4a_0$ and $6a_0$ but higher for the simulation cell of length $L = 8a_0$. Thus, it can be expected that in such a ZGNR, spins in domains of the unreconstructed zigzag edge between 57 pairs should change their orientation once these domains become smaller than 6 hexagons in length. For the wider ZGNRs, the ferromagnetic final state is lower energy for all the considered simulation cells $L = (4 - 8)a_0$. Therefore, the transition to the ferromagnetic state should occur at even larger lengths of domains of the unreconstructed edge separating 57 pairs or lower densities of 57 pairs.

In the limit of isolated 57 pairs, the magnetic state of the ZGNR with a 57 pair should be the same as in the perfect ZGNR. To estimate the reaction energy for an isolated 57 pair, it makes sense, therefore, to consider the final state with the antiferromagnetic ordering of spins at the opposite ZGNR edges. However, performing the calculations for insufficiently large simulation cells, the system can get into the more energetically favourable ferromagnetic state. Or even in sufficiently large simulation cells, the system can converge to the energetically unfavourable but metastable ferromagnetic or nonmagnetic states providing an additional error in the reaction energy of hundreds meV. Therefore, for accurate calculations of energetics of the reactions at graphene edge, care should be taken to converge to the proper magnetic state.

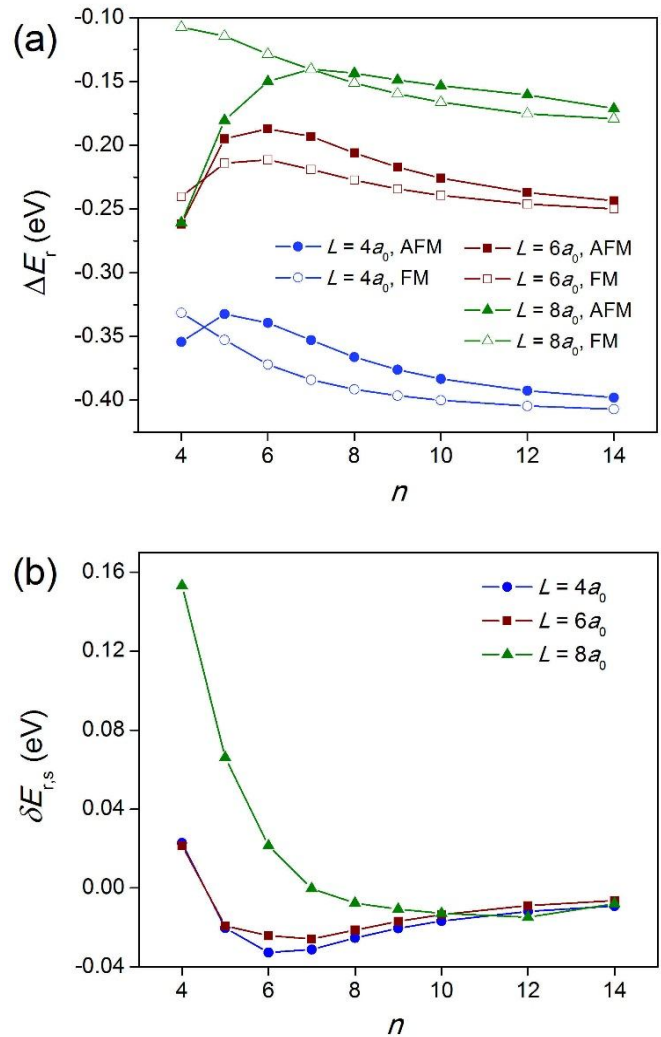

**Fig. 6.** (a) Calculated reaction energies $\Delta E_r$ (in eV) for the formation of a pentagon-heptagon (57) pair at the unreconstructed zigzag edge of a fully free and carbon-terminated zigzag graphene nanoribbon as functions of the number $n$ of zigzag rows for the final states of the nanoribbon with a 57 pair with the antiferromagnetic (AFM, closed symbols and solid lines) and ferromagnetic (FM, open symbols and dashed lines) ordering of spins at the opposite nanoribbon edges. (b) Calculated difference $\delta E_{r,s}$

(in eV) of the reaction energies for the final states with the ferromagnetic and antiferromagnetic spin ordering as a function of the number $n$ of zigzag rows. The lengths $L$ of the simulation cell along the nanoribbon axis are $4a_0$ (blue circles), $6a_0$ (red squares) and $8a_0$ (green triangles), where $a_0$ is the lattice constant of graphene.

The results presented in Fig. 6a already demonstrate that the reaction energy $\Delta E_r$ depends strongly on the length $L$ of the simulation cell in the range from $4a_0$ to $8a_0$. The reaction energy increases by 0.09 – 0.16 eV for the considered ZGNRs of different width and the same magnetic state of the ZGNR with a 57 pair upon increasing the length of the simulation cell from $4a_0$ to $6a_0$. A further increase by up to 0.13 eV (up to 0.07 eV for the final state with the antiferromagnetic ordering of spins at the opposite ZGNR edges) is observed upon increasing the length of the simulation cell to $8a_0$. Fig. 7 shows how the reaction energy $\Delta E_r$ and activation barrier $E_a$ for formation of the first 57 pair converge upon increasing the length $L$ of the simulation cell for the 4- and 6-ZGNRs in the simulation cells of length $L = (4 - 14)a_0$. It is seen that when the ZGNR with a 57 pair is assumed to be in the antiferromagnetic state, the reaction energy $\Delta E_r$ and activation barrier $E_a$ for both of the ZGNRs reach the plateau at $L = 8a_0$. The deviations of the reaction energy and activation barrier for $L = 8a_0$ from the data obtained for the larger simulations cells lie within 0.05 eV. For $L = 6a_0$, the activation barrier for the 4-ZGNR is underestimated by more than 0.1 eV and the reaction energy for the 6-ZGNR is also slightly below the value at the plateau. Therefore, it can be stated that for adequate modeling of defects at the zigzag edge, such as 57 pairs, the periodic images of the defects should be separated by at least 6 hexagons. It should be noted that the reaction energy for the final state of the ZGNR with a 57 pair with the ferromagnetic spin ordering keeps increasing upon increasing the length of the simulation cell (Fig. 7). In the limit of domains of the unreconstructed zigzag edge separating the 57 pairs, it can be expected that the relative energy of the ferromagnetic and antiferromagnetic states is proportional to the length of these domains.

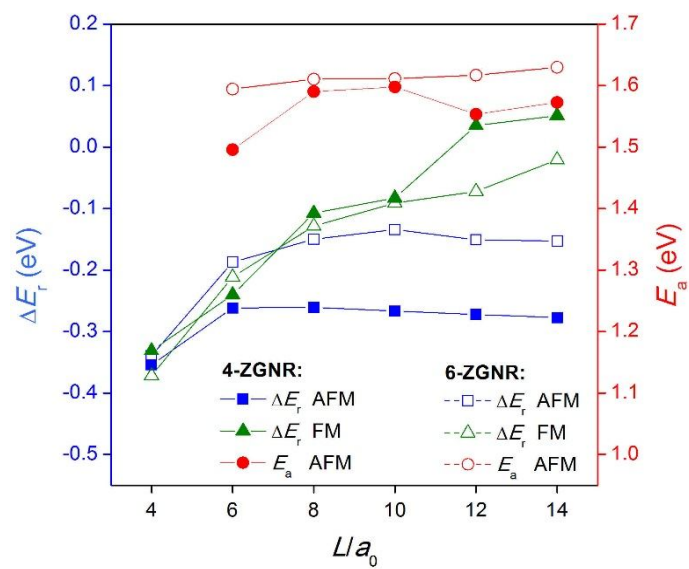

**Fig. 7.** Calculated energetic characteristics of the formation of pentagon-heptagon (57) pair at the unreconstructed zigzag edge of a fully free and carbon-terminated zigzag graphene nanoribbon as functions of the length $L$ of the simulation cell along the nanoribbon axis divided by the lattice constant of graphene $a_0$: reaction energies $\Delta E_r$ (in eV, left axis) for the final states of the nanoribbon with a 57 pair with the antiferromagnetic (AFM, blue squares) and ferromagnetic (FM, green triangles) ordering of spins at the opposite nanoribbon edges and activation barrier $E_a$ (in eV, red circles, right axis) for the final state with the antiferromagnetic spin ordering. The results for the nanoribbons consisting of 4 and 6 zigzag rows are shown by closed and open symbols, respectively.

Based on the calculations for the 6-ZGNR in the simulation cell of length $L = 8a_0$ (Figs. 6a and 7), we conclude that the converged values of the reaction energy $\Delta E_r$ and activation barrier $E_a$ are about –0.15 eV and 1.61 eV, respectively. Thus, although the system energy decreases upon formation of the first 57 pair, the reaction should be very slow at room temperature. This can explain why the reconstructed zigzag edges are not abundant in the experiments [38–42] and high temperature [42] or electron impacts [38,40] are needed for the reconstruction to take place. Our results agree with the data from Ref. [22], where the reaction energies from –0.18 eV to –0.24 eV were reported for ZGNRs consisting from 6 to 12 zigzag rows for the simulation cell of length $L = 8a_0$ and the activation barrier was also 1.61 eV. Different activation barriers of 1.12 eV and 1.80 eV and reaction energies of almost -1 eV and +0.2 eV were obtained in Refs. [12] and [13], respectively. In those papers, the authors considered narrower 4-ZGNRs with one zigzag row fixed at the edge opposite to the one where the 57 pair was formed. Furthermore, the simulation cells were only of lengths $L = 4a_0$ and $6a_0$, respectively. Therefore, the deviations in the results of Refs. [12] and [13] can be attributed to the insufficient ZGNR width and, most importantly, the insufficient length of the simulation cell considered in those papers.

Let us also briefly discuss the calculations for simulation cells that are shorter than $L = 8a_0$ in length for which the convergence is not yet achieved (Figs. 6a and 7). These results correspond to the high density of 57 pairs. They indicate that the presence of other 57 pairs nearby should lead to a decrease of the reaction energy and activation barrier. This is consistent with the data from literature for simultaneous reconstruction of the whole zigzag edge for which the barrier of 0.6 eV per hexagon pair was obtained [21,46]. Also the calculations for sequential generation of 57 pairs showed a decrease of the activation barrier at each step [13,48]. Therefore, it is likely that once a 57 pair is formed, new 57 pairs are generated close to it. Such a conclusion is in agreement with the experimental observations [38,40]. A large simulation cell should be considered to model properly generation of consecutive 57 pairs under periodic boundary conditions excluding the interaction of periodic images. This study will be performed elsewhere.

## 4. CONCLUSIONS

In the present paper, we studied how the parameters of the ZGNR used as a model of a semi-infinite graphene layer affect the results of DFT calculations of reactions at graphene edges by the example of the first stage of reconstruction of zigzag graphene edges, formation of the first 57 pair. It was found that the relative energy of the fully reconstructed zigzag edge with respect to the unreconstructed one converges fast upon increasing the ZGNR width. The width of 4 zigzag rows is sufficient to reproduce this quantity with the accuracy of 0.01 eV/Å. Such computational tricks as the termination of the edge opposite to the one where the reconstruction takes place by hydrogen atoms or fixation of atoms at that edge in addition to being unrealistic, do not speed up the convergence. Therefore, free ZGNRs with no hydrogen termination were used to study the formation of the first 57 pair.

To investigate how the reaction energy and activation barrier for the formation of the first 57 pair at the zigzag edge change upon increasing the ZGNR width and decreasing the defect density, ZGNRs of different width in simulation cells of different length along the ZGNR axis were considered. We found that convergence of the reaction energy and activation barrier for formation of the first 57 pair within 0.06 eV, i.e. close to the chemical accuracy, is achieved for the ZGNR width of 6 zigzag rows and length of the simulation cell of 8 elementary periods (about 20 Å). The latter result means that periodic images of 57 pairs should be separated by at least 6 hexagons (about 15 Å). Such requirements on the model GNRs should be taken into account in further studies of generation of subsequent 57 pairs for consecutive zigzag edge reconstruction and simulations of other phenomena at graphene edges such as edge etching [42,44,54,55], formation of atomic chains [38,56,57], migration of atoms [22,37,53], graphene growth [58–60], chemical absorption [14,21,25,27,29,31–36], etc.

The converged values of the reaction energy and activation barrier are –0.15 eV and 1.61 eV, respectively. The high value of the activation barrier explains why the reconstructed edges are not that often seen in the experiments [38–42] but can be formed upon electron irradiation [38,40] or heating [42]. The obtained dependences on the length of the simulation cell indicate that the reaction energy and activation barrier should decrease upon increasing the defect density, i.e. it is likely that new 57 pairs are generated close to the ones already formed [38,40].

Our calculations also demonstrated that the magnetization of the zigzag edge gets reduced only locally upon formation of an isolated 57 pair. Because of the local effect of the 57 pair on the edge magnetization, we expect that such defects should not influence much the spin transport in graphene nanoribbons as long as their density is low. On the other hand, we showed that upon decreasing the distance between 57 pairs, spins in domains of the unreconstructed zigzag edge separating the 57 pairs should change their orientation. For the 6-ZGNR, the ferromagnetic ordering of spins at the opposite ZGNR edges becomes more energetically favourable than the antiferromagnetic one once the 57 pairs are separated by less than 6 hexagons. For wider ZGNRs, the transition to the ferromagnetic state is expected even at a larger separation of the 57 pairs. This phenomenon can be used to tune the ZGNR performance in spintronic devices [30,48,61,62].

## ACKNOWLEDGEMENTS

IVL acknowledges the European Union MaX Center of Excellence (EU-H2020 Grant No. 824143). This work has been carried out using computing resources of the federal collective usage center Complex for Simulation and Data Processing for Mega-science Facilities at the NRC ''Kurchatov Institute''.

## DATA AVAILABILITY STATEMENT

The raw data required to reproduce these findings are available to download from https://data.mendeley.com/datasets/v9vf78f6rv/1.